\documentclass[traditabstract]{aa} % for the abstract without structuration
\usepackage{graphicx}
\usepackage{txfonts}
\begin{document}
\title{The upper atmosphere of the exoplanet HD\,209458\,b revealed by the sodium D lines}
\subtitle{Temperature--pressure profile, ionization layer, and
thermosphere}

 % \title{The upper atmosphere of the exoplanet HD209458b from the sodium D lines. T--P
 % profile, ionization layer and thermosphere.
 % }
%   \titlerunning{Ionosphere and thermosphere detected\hd}

  \author{A. Vidal--Madjar\inst{1}
  \and D. K. Sing\inst{2}
  \and A. Lecavelier~des~Etangs\inst{1}
  \and R. Ferlet\inst{1}
  \and J.--M. D\'esert\inst{3} \and G.~H\'ebrard\inst{1} \and I. Boisse\inst{1}
  \and D.~Ehrenreich\inst{4} \and C. Moutou\inst{5} }

  \institute{
    CNRS, UPMC,
    Institut d'astrophysique de Paris, UMR 7095,
    98$^{\rm bis}$ boulevard Arago 75014 Paris, France, \email{alfred@iap.fr}
         \and
    Astrophysics Group, School of Physics, University of Exeter, Stocker Road, Exeter EX4 4QL, UK
        \and
    Harvard--Smithsonian Center for Astrophysics, 60 Garden St., Cambridge, MA 02138, USA
        \and
        Laboratoire d'Astrophysique de Grenoble, Universit\'e Joseph Fourier,
        CNRS (UMR 5571), BP 53, 38041 Grenoble cedex 9, France
        \and
    Laboratoire d'Astrophysique de Marseille, Technop\^ole Marseille \'Etoile,
    38 rue Fr\'ed\'eric Joliot Curie, 13013 Marseille, France
  }

 \date{Received ...; accepted ...}

%\abstract{}{}{}{}{}
% 5 {} token are mandatory

\begin{abstract}
% \abstract
{ A complete reassessment of the \emph{Hubble Space Telescope}
(HST) observations of the transits of the extrasolar planet
HD\,209458\,b has provided a transmission spectrum of the
atmosphere over a wide range of wavelengths. Analysis of the NaI
absorption line profile has already shown that the sodium
abundance has to drop by at least a factor of ten above a critical
altitude. Here we analyze the profile in the deep core of the NaI
doublet line from HST and high--resolution ground--based spectra
to further constrain the vertical structure of the HD\,209458\,b
atmosphere.

    With a wavelength--dependent cross section that spans more than 5 orders of magnitude,
    we use the absorption signature of the NaI doublet as an atmospheric probe.
      The NaI transmission features are shown to sample the atmosphere of
    HD\,209458\,b over an altitude range of more than 6\,500 km, corresponding to
    a pressure range of 14 scale heights spanning 1 millibar to $10^{-9}$ bar pressures.
By comparing the observations with a multi--layer model in which
temperature is a free parameter at the resolution of the
atmospheric scale height, we constrain the temperature vertical
profile and variations in the Na abundance in the upper part of
the atmosphere of HD\,209458\,b.

We find a rise in temperature above the drop in sodium abundance
at the 3~mbar level. We also identify an isothermal atmospheric
layer at $1\,500\pm100$ K spanning almost 6~scale heights in
altitude, from 10$^{-5}$ to 10$^{-7}$ bar. Above this layer, the
temperature rises again to $2\,500^{+1\,500}_{-1\,000}$ K at
$\sim$10$^{-9}$ bar, indicating the presence of a thermosphere.

 The resulting temperature--pressure (T--P) profile agrees with the Na condensation
  scenario at the 3 mbar level, with a possible signature of sodium
  ionization at higher altitudes, near the 3$\times$10$^{-5}$ bar level.
  Our T--P profile is found to be in good agreement with the profiles obtained with
  aeronomical models including hydrodynamic escape.
}

\end{abstract}


\titlerunning{HD\,209458\,b T--P profile from the sodium D lines}
\authorrunning{Vidal--Madjar et al.}

  \maketitle
%
%________________________________________________________________

\section{Introduction}
\label{sec:intro}

Only a few detections of extrasolar planets' atmospheric species
are reported so far, but they have been recognized as important
steps in our understanding of these objects. Of particular
interest are {\sl Hubble Space Telescope} (HST) observations of
HD\,209458\,b during primary transit, which yielded the detection
of NaI (\cite{Charbonneau02}), as well as HI, OI, and CII
(\cite{Madjar03,Madjar04}), HI (\cite{Ballester07}), Rayleigh
scattering by H$_2$ (\cite{Lecavelier08b}), along with upper limits
on the presence of TiO and VO (\cite{Desert08}) and recently
confirmation of CII and detection of SiIII (\cite{Linsky10}) and
even possibly SiIV (\cite{Schlawin10}). HD\,189733\,b is the second
transiting planet for which HST observations yielded information
on the atmospheric transmission spectrum. HST/NICMOS has been used
to detect H$_2$O and CH$_4$ (Swain et al. 2008), though the H$_2$O
feature has been challenged by a more sensitive search using
filter photometry (\cite{Sing09,Desert09,Desert10}) showing the
presence of haze condensate at high altitude as previously
observed in the optical using HST/ACS (\cite{Pont08,
Lecavelier08a}). In addition to HST, ground--based,
high--resolution spectra have also allowed sodium detection in
HD\,189733\,b (\cite{Redfield08}), confirmed sodium in HD\,209458\,b
(\cite{Snellen08}), detected CO and even winds in the planetary
atmosphere (\cite{Snellen10}), while the other important alkali
metal potassium has been detected in XO--2b using narrowband
photometry (\cite{Sing10}).

Sing et~al. (2008a) revisited all the existing HST/STIS
spectroscopic data of HD\,209458\,b transits gathered over the
years. Their result is an estimate of the transit absorption depth
(AD) as a function of wavelength from 3\,000~\AA\ to 8\,000~\AA .
The variation of absorption depth as function of wavelength has
been interpreted by atmospheric signatures of Na, H$_2$ Rayleigh
scattering and TiO--VO molecules
(\cite{Sing08b,Lecavelier08b,Desert08}).
%(Sing et~al. 2008b;
%Lecavelieret~al. 2008b; D\'esert et~al. 2008a)
Another important result obtained with the analysis of this
transit absorption spectrum is the observation of a sharp drop in
sodium abundance above the altitude corresponding to the
absorption depth of 1.485\% (\cite{Sing08a,Sing08b}). This drop is
clearly seen in the profile of the sodium absorption line, which
differs from the classical ``Eiffel Tower'' shape found when there
is a constant sodium abundance (\emph{e.g.}, Seager \& Sasselov
2000). The sodium line shows a plateau with an almost constant
absorption depth around AD$\sim$1.485\% at several hundred
Angstroms around the line center. Closer to the line center in the
line core, the absorption depth increases again with a shape more
like the ``Eiffel Tower'' above the plateau, which shows that the
sodium is still present above the altitude of the abundance drop.
To match the width of the core of the sodium line, the drop in
sodium abundance should be at least a factor of 10
(\cite{Sing08a,Sing08b}). Moreover, if the interpretation of the AD
rise at wavelengths shorter than $\sim$~5\,000~\AA\ in terms of
Rayleigh scattering by H$_2$ molecules is correct, then the
pressure at the altitude of the abundance drop is constrained to
be about 3~mbar (\cite{Lecavelier08b}).

\begin{figure}
\includegraphics[angle=90,width=\columnwidth]{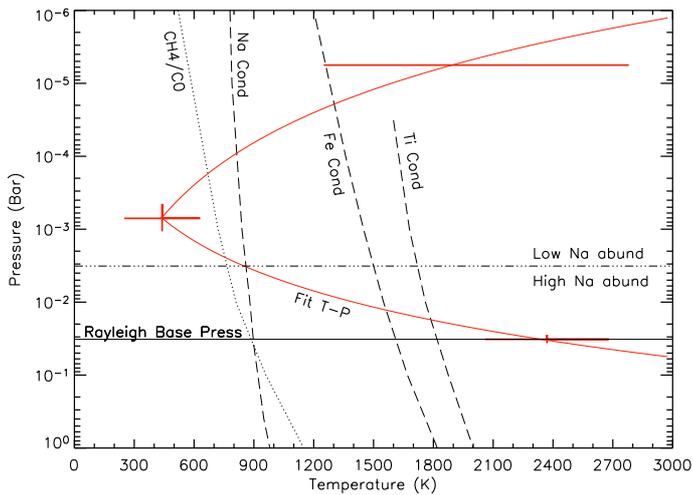}
\caption{The parametric T--P profile of the atmosphere of
HD\,209458\,b (thick red solid line), as adjusted over HST/STIS
observations by Sing et~al. (2008b). The three adjusted (T,P)
parameters are shown with their temperature error bars (red). The
tripled--dot--dashed line shows the level of Na abundance drop at
3~mbar pressure. The thin solid line shows the 32~mbar pressure
measured at the base of the sodium absorption line using the
Rayleigh scattering. The dotted line shows the CH$_4$/CO
equilibrium curve. The dashed lines show the condensation curves
for sodium, iron, and titanium.} \label{fig:TPprofile}
\end{figure}

The fit of the HD\,209458\,b absorption spectrum including
profiles of sodium and Rayleigh scattering by H$_2$ yielded (Sing
et~al. 2008b) an estimate of the temperature--pressure (T--P)
profiles for the atmosphere from $\sim$30~mbar to
$\sim$10$~\mu$bar (see also Madhusudhan \& Seager (2009) for
related issues when retrieving temperature and abundances from
transit data). This result has been obtained using a parametric
profile to fit the data, assuming linear variations in temperature
with altitude and two layers with constant slope of temperature
versus altitude (Fig.~\ref{fig:TPprofile}). Nonetheless, with this
method, two different profiles provide an equally good fit to the
absorption spectrum: the first profile involves Na condensation to
explain the abundance drop at the absorption depth of 1.485\%; in
the second profile the temperature is not low enough to allow the
sodium condensation, and the decrease in sodium absorption at
altitude above the plateau is interpreted by ionization. The two
resulting T--P profiles are

\begin{enumerate}

\item In the first profile the temperature decreases down to less
than 900~K, causing a sodium abundance drop by \emph{condensation}
at about 3~mbar (Fig.~\ref{fig:TPprofile});

\item The second profile (not shown in Fig.~\ref{fig:TPprofile})
is nearly isothermal at about 2200~K, in which case the abundance
drop is explained by partial \emph{ionization} of atomic sodium
above the 3~mbar level in a scenario proposed by Fortney et~al.
(2003).

\end{enumerate}

Both profiles do not present variations similar to model
predictions as calculated in the literature specifically for
HD\,209458\,b. The direct comparison between extracted
observational T--P profiles and model predictions will be made in
Sect. 4 (Fig.~\ref{fig:T-P}).

To further reduce the impact of a priori hypothesis on the data
interpretation, we propose here a new analysis of the absorption
depth profile with particular attention to the inner part of the
NaI doublet. This new analysis is done using a non--parametric
approach, with no a priori assumption about any expected behaviour
related to the T--P profile or to supposed NaI abundance
variations. The hope is to be able to distinguish between the two
scenarios described above, avoiding a priori assumptions. In
Sect.~\ref{sec:re} we present the method for the data analysis. In
Sect.~\ref{sec:variations} we compare the observations to the
model calculations. The results are discussed in
Sect.~\ref{sec:discuss} where a new description of the upper
atmospheric layers of HD\,209458\,b is proposed.

\section{The sodium absorption depth in the line core}
\label{sec:re}

\begin{figure}
\includegraphics[width=0.48\textwidth,angle=0]{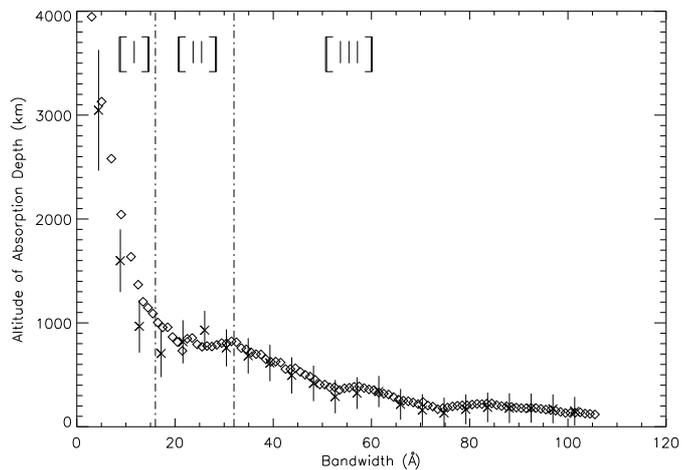}%{Obs_vs_altitude_Ref0.0149145_v2.eps}
\caption{The relative altitude of the absorption as observed by
Sing et~al. (2008a) close to the NaI lines in variable bandwidths
(see text). Their photometric approach is presented as crosses
with the corresponding error bars, while their spectroscopic
approach is plotted as diamonds. The error bars corresponding to
both evaluations are similar. These observations are correlated
since extracted over variable bandwidths all centered on the NaI
lines. Three main regions could be seen: [I] corresponding to a
sharp decrease, [II] to a plateau and [III] to a more gentle
decrease. Analysis of these three regions will be carried out
starting from region [III] inward.} \label{fig:obs1}
\end{figure}

Sing et~al. (2008a) evaluated the transit absorption depth around
the center of the NaI lines in variable bandwidths from 4.4 to
100\AA\ centered over both lines. At 12\AA\ bandwidth, the
measured absorption depth agrees with the Charbonneau et~al.
(2002) value. In the Sing et~al. (2008a) approach, the absorption
depth is evaluated \emph{relatively} to a reference absorption
depth in a fixed double spectral band (noted \emph{Ref}) on both
sides of the NaI lines: from 5818 to 5843\AA\ and from 5943 to
5968\AA , \emph{i.e.} in a fixed 50\AA\ region. The measurements
are obtained by first integrating the raw data over the selected
bandwidth before extraction of the absorption depth.
%This reduces systematics due to the stellar limb darkening
%and the intra--orbit HST thermal focus variations {\bf (??)}.
Because the measurements correspond to the total absorption depths
within various bands, including the line center as done with
filters, this absorption spectrum as a function of the bandwidth
can thus be regarded as a spectrum obtained with a ``photometric''
approach. The resulting ``photometric'' spectrum is plotted in
Fig.~\ref{fig:obs1}.

In another approach, Sing et~al. (2008a) also extracted
\emph{spectroscopically} the absorption depth distribution over
each pixel in the whole spectral range covered by the
observations. These spectroscopic absorption depths can also be
averaged over different spectral bandwidths as in the photometric
approach. We calculated the difference between the spectrum
averaged over various bands centered on the NaI lines and the
spectrum averaged in the reference band, \emph{Ref}, defined
above.
%similar to the one used in the photometric approach.
The result obtained in this ``spectroscopic'' approach (see
Figure~\ref{fig:obs1}) can be directly compared with
``photometric'' result. Small differences do show up between the
two extraction methods, but they are well within the error bars.
The main differences probably arise from the limb--darkening
correction, which is applied at each pixel wavelength in the
spectroscopic approach, while the correction is a weighted average
limb--darkening correction in the considered bands in the
``photometric'' approach. Therefore, the ``spectroscopic'' is
expected to provide a better correction for the limb--darkening
variations within the stellar lines themselves and in particular
within the two strong NaI doublet lines. In the following, we use
the spectroscopic absorption depths.

In Figure~\ref{fig:obs1} the measured absorption depths have been
translated into planetary radii, i.e., altitudes above the
\emph{Ref} level, defined by the level of the absorption depth in
the \emph{Ref} spectral domain. To translate AD into altitudes
above the \emph{Ref} level, we used R$_{\rm P}$=1.32R$_{\rm Jup}$
(\cite{Knutson07}) for the planetary radius and an absorption depth
of AD$_{Ref}$~=~1.49145\%\ at the \emph{Ref} level. As expected,
the observed altitudes (\emph{i.e.} AD) vary as a function of the
bandwidth with a slow rise toward smaller bandwidths; however,
three different regions could be seen in terms of AD variations: a
region [I] (see Fig.~\ref{fig:obs1}) presenting a rapid decrease
of the AD with increasing bandwidths followed by a kind of plateau
(region~[II]), a puzzling feature discussed in
Sect.~\ref{sec:discuss} followed by a last region [III] where a
slow decrease in the AD could be seen. These observations are not
independent. In particular, the error bars are larger than the
observations' fluctuations. Indeed, all observations made in a
given bandwidth are present in observation--extracted in any
larger bandwidth. For that reason, we will complete our analysis
of the observed behavior of the AD by starting from the larger
bandwidths and moving inward. This approach will give us first
access to the atmospheric information at the lower altitudes and
then, by eliminating this information in the narrower bandwidths,
move progressively upwards in the atmosphere.

\section{Modeling the absorption spectrum}
\label{sec:variations}

The variations in the absorption depth (or variations of the
corresponding altitude) along the line profile are due to
variations of the line opacity. At each wavelength, the absorption
line probes different altitudes in the planetary atmosphere with a
resolution on the order of the error bar on the absorption
estimate, \emph{i.e.} about 0.01\%\ in absorption depth
corresponding to about 315~km in altitude. Because the absorption
cross section increases toward wavelengths closer to the core of
the lines, the absorption depth increases toward narrower
bandwidths (Fig.~\ref{fig:obs1}). These variations in absorption
altitude (or absorption depth) can be interpreted through
comparison with model calculations.

\subsection{From absorption depth to physical quantities}

To interpret the measured variations in absorption with
bandwidths, we built the simplest possible model, \emph{i.e.} an
isothermal hydrostatic uniform abundance model (\emph{IHUA}). The
model with isothermal hydrostatic atmosphere at temperature $T$
with a single sodium abundance value ([Na/H], relative to the
solar abundance), and a pressure $P_{z=0}$ at a reference level,
is a three--parameter model ($T$,[Na/H],$P_{z=0}$). This model is
aimed at a better understanding of the observed absorption depth
\emph{variations} in terms of departures from the \emph{local}
\emph{IHUA} conditions, \emph{i.e.} in terms of variations in
temperature or sodium abundance, without any {\it a priori}
hypothesis on the shape of the vertical temperature--pressure
profile as needed in Sing et~al. (2008a). Despite its simplicity,
assuming hydrostatic equilibrium, this (\emph{IHUA}) model still
holds \emph{locally} to interpret observational measurements
because the atmospheric scale height is on the order of the
observational accuracy, \emph{i.e.} $\sim$~300~km for typical
temperatures in the HD\,209458\,b atmosphere. The exponential
decrease in the volume density with altitude is very steep; as a
result, the variations in the measured absorption at a given
altitude are directly related to the physical conditions
(temperature, abundance) in the atmosphere at the corresponding
altitude.

The model absorption depth is calculated in a straightforward
manner following Lecavelier Des Etangs et~al. (2008a,b). The total
column density along the line of sight passing through the
terminator at different altitudes is given by Fortney et~al.
(2005). The optical depth, $\tau$, in a line of sight grazing the
planetary limb at an altitude $z$ is given by
\begin{equation}
\tau(\lambda,z)\approx \sigma(\lambda)n(z)\sqrt{2\pi R_{P} H}
\label{tau_lambda}
\end{equation}
where $R_{P}$ is the planetary radius, $H$ the atmosphere scale
height, and $n(z)=n_{(z=0)} exp(-z/H)$ the volume density at the
altitude $z$ of the main absorbent with a cross section
$\sigma(\lambda)$. The scale height is given by the relation
$H=kT/\mu g$ where $k$ is the Boltzmann constant, $T$ the
temperature, $\mu$ the mean mass of atmospheric particles taken to
be 2.3 times the mass of the proton, and $g$ the gravity at the
planetary radius, $g = M_{P}{\rm G}/R_{P}^2$ with $M_{P}$ equal to
the planetary mass and G the gravitational constant. The
absorption at a given wavelength, $\lambda$ is calculated by
finding the altitude at which the optical thickness is
$\tau(\lambda,z)=\tau_{eff}$, where $\tau_{eff}$=0.56
(\cite{Lecavelier08a}).

Using the equations and quantities defined above,
the effective altitude $z$ is given by
\begin{equation}
z(\lambda)=H\ln \left(\frac{{\rm [Na/H]}\times P_{z=0}\times
\sigma_{NaI}(\lambda)} {\tau_{eff}}\times\sqrt{2\pi  R_P/kT \mu g}\right)
\label{z_lambda}
\end{equation}
where $\sigma_{NaI}$ is the NaI lines cross section. Equation
\ref{z_lambda} shows that the altitude (or corresponding
absorption depth) depends on pressure and abundance, but the
variation in the altitude as a function of wavelength does not
depend upon pressure or abundance, but only upon the scale height,
which is directly and only related to $T$, the temperature at the
corresponding altitude. Therefore, the interpretation of
Fig.~\ref{fig:obs1} (in particular the slope of the absorption
depth curve as a function of wavelength) provides the temperature
profile in the altitude range probed by the sodium line core: from
100 to 4\,000~km altitude. This temperature profile can have a
typical resolution close to the atmospheric scale height.

As shown by Eq.~\ref{z_lambda}, the absolute values of the
pressure and abundance impact only the absolute absorption
altitudes and not their variations. Moreover, there is a
degeneracy between these two quantities because only the product
$[Na/H]\times P_{z=0}$ can be constrained by the measurement of
the absorption altitude. In other words, a change in the abundance
can be directly compensated by a similar change in the pressure.
That is the total quantity of gas in the line of sight.

\subsection{Pressure at the reference level}
\label{Ref_Pressure}

 One could assume that the pressure at the reference level,
$P_{z=0}$ for our NaI study in the core of the lines, {\it i.e.}
as evaluated over the reference domain $Ref$ over which the
AD$_{Ref}$ evaluation is made and relative to which all our
differential AD evaluations are made, is given by the H$_2$
Rayleigh scattering observed at wavelengths shorter than
$\sim$5\,000~\AA\ (\cite{Lecavelier08b}). This leads to a reference
pressure $P_{z=0}\sim$3~mbar as shown in Fig.~\ref{fig:TPprofile}.
This pressure corresponds to the location where the NaI abundance
drops and above which only the core of the NaI lines do give an
absorption signature.

To be more precise, one could alternatively use the evaluated NaI
abundance, $[Na/H]\sim 2$ found in the lower layers of the
atmosphere by Sing et~al. (2008a), to evaluate the corresponding
pressure at the reference altitude. Indeed with only the product
$[Na/H]\times P_{z=0}$ being defined, by using the evaluated NaI
abundance in the lower layers one can deduce the pressure level
sampled by the NaI lines at any spectral region. This is given by
the following relation deduced from Eqs.~\ref{tau_lambda}
and~\ref{z_lambda}.

\begin{equation}
P(\lambda)= \left(\frac {\tau_{eff}\times kT} {{\rm
[Na/H]}\times\sigma_{NaI}(\lambda) \times \sqrt{2\pi R_P/kT \mu
g}}\right). \label{P_lambda}
\end{equation}

\begin{figure}
\includegraphics[width=0.48\textwidth,angle=0]{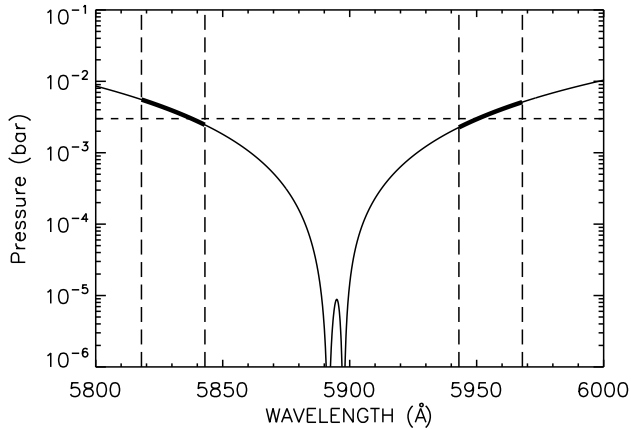}
\caption{The variation in the pressure as a function of wavelength
is given according to Eq.~\ref{P_lambda} (solid line). Twice the
solar NaI abundance is assumed. A temperature of 900K
(corresponding to the sodium condensation temperature) is also
assumed and the two spectral reference domains are indicated as
vertical long dashed lines. In those regions the evaluated
pressure (thick solid lines) is indeed on the order of 3~mbar as
indicated by the horizontal dashed line.} \label{fig:Ref_pressure}
\end{figure}

The result of Eq.~\ref{P_lambda} is shown in
Fig.~\ref{fig:Ref_pressure} for which the assumed abundance is
twice the solar abundance, valid up to our reference level. The
pressure broadening in the NaI line wings is larger than the
natural width broadening at pressures above 29~mbar as it could be
scaled from the corresponding values at 1~bar and 2\,000~K where
the inverse lifetime related to pressure broadening is $\tau =
0.071$cm$^{-1}\times c = 2.1\times 10^{9}$s$^{-1}$ to be compared
to the natural inverse lifetime which is $6.1\times
10^{7}$s$^{-1}$ (\cite{Iro05}). Because here we are analyzing the
inner parts of the NaI lines (between the two reference domains,
see Fig.~\ref{fig:Ref_pressure}), we only investigate the upper
part of the atmosphere where lower pressures are present. We thus
can ignore here the pressure broadening effects. Consequently,
$\sigma_{NaI}(\lambda)$ in Eq.~\ref{P_lambda} was evaluated by
taking only into account the Doppler core broadening and the
natural wing broadening of both NaI doublet lines.

We find that, using a temperature of 900~K and sodium abundance
twice solar, absorption over the two spectral reference domains,
on both sides of the NaI lines from 5\,818 to 5\,843~\AA\ and from
5\,943 to 5\,968~\AA , is due to sodium located at a pressure of 3
mbar (Fig.~\ref{fig:Ref_pressure}).

If a different temperature is selected for the reference level (as
for instance 2\,000~K as suggested in the second possible
scenario), then the evaluated pressure at the reference level
would be slightly higher and on the order of 6~mbar, a value still
within the pressure domain where the pressure broadening effect
could be ignored. With this new reference pressure, the T--P
profile obtained here would still hold with only a shift in the
reference level from 3 to 6~mbar.

Once the reference pressure has been set, our \emph{IHUA} model
becomes a two--parameter model ($T$,[Na/H]) and the selection of a
different reference pressure only slightly shifts our temperature
evaluations along the pressure scale, with the sodium abundance
modified by the same ratio.

\begin{figure}
\includegraphics[angle=0,width=9.cm]{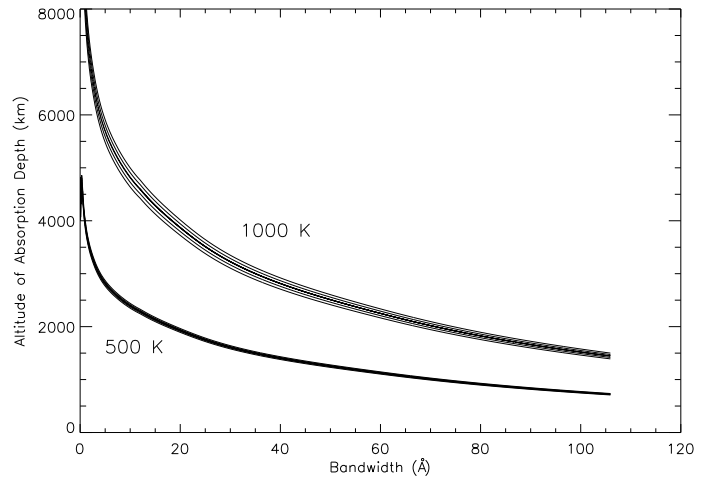}%{Variations_AD_T2000_Ref-var_Na-var.eps}
\caption{Absorption depths calculated using \emph{IHUA} models
with $T$=500~K and $T$=1\,000~K  for various sodium abundances
ranging from [Na/H]=10$^{-2}$, 10$^{-4}$, 10$^{-6}$, 10$^{-8}$,
10$^{-10}$ (top to bottom).
%The slight differences are due to the different weights of the averages
%since more
%or less slope in the absorption profiles over the different
%bandwidth domains considered do show up with a changing [Na/H]
%abundance.
} \label{fig:1}
\end{figure}
%\end{figure*}

\subsection{Temperature profile}

In the framework of our study, we consider \emph{relative}
absorption depths, calculated by the difference in absorption
depths between two spectral domains: the first one for a given
bandwidth centered on the sodium lines, the second one in the
constant \emph {Ref} domain. If the [Na/H] abundance varies, the
absolute absorption depth of all spectral domains changes by
nearly the same amount (see Eq.~\ref{z_lambda}). As a result, the
\emph{relative} absorption depth in the sodium line does not
depend on the sodium abundance as shown in Figure~\ref{fig:1}.
Indeed a change in the abundance produces about the same shift in
the absorption depth at all wavelengths. Therefore, the difference
between the absorption depth in a given bandwidth and the
absorption depth in the reference wavelength domain
(AD--AD$_{Ref}$) barely depends on the sodium abundance. On the
contrary, the slope of the models clearly changes with
temperature. The sodium abundance thus cannot be estimated by the
transmission spectrum of \emph{relative} absorption depth.

However, the variations in the \emph {relative} absorption depth
does allow estimating the local temperature $T$ at the absorption
altitude. In effect, the variation in the absorption altitude with
wavelength depends on the variation in the horizontal integrated
density with altitude. For instance, if one considers two
wavelengths $\lambda_1$ and $\lambda_2$, with corresponding line
cross section $\sigma(\lambda_1)> \sigma(\lambda_2)$, a grazing
line of sight will have the same optical thickness at these two
wavelengths if they are at altitude $z_1$ at $\lambda_1$ and $z_2$
at $\lambda_2$ such that the variation in partial density of the
absorbing element $n(z)$ between $z_1$ and $z_2$ compensates for
the different cross sections. The optical thickness at these two
wavelengths are equal if
$n(z_1)\sigma(\lambda_1)=n(z_2)\sigma(\lambda_2)$. Therefore, the
variation in the observed absorption altitude as a function of
wavelength (equivalent to variation as a function of line cross
section) reveals the variation in the partial density at the
considered altitude. At hydrostatic equilibrium, the density
variation with altitude is determined by the temperature through
the characteristic scale height. As a consequence, the temperature
constrains the slope of the absorption altitude as a function of
wavelength. The higher the temperature, the larger the slope. For
a given absorption feature with a known cross section as a
function of wavelength, the temperature $T$ can be derived from
the measurements of the absorption altitude spectrum using
(\cite{Lecavelier08a}):

\begin{equation}
T=\frac{\mu g}{k} \left( \frac{d\ln \sigma}{d\lambda}\right)^{-1}
\frac{d z(\lambda)}{d\lambda}. \label{T_dz}
\label{Eq:T}
\end{equation}

%A final comment regarding the way we use the \emph {IHUA} models.
%We are indeed not sensitive to the global Na abundance in this
%approach because the reference absorption depth, AD$_{Ref}$, is caused by
%another absorber in the atmosphere, independent of the one we are
%considering in our \emph {IHUA} model calculations.  More precisely,
%in our study AD$_{Ref}$ is determined by the lower and denser
%layers where the Na abundance was demonstrated to be much higher. Above
%these layers, we have thus simply to correct the calculated \emph
%{variable} AD$_{Ref}$ value (which depends upon both $T$ and
%[Na/H]) by the observed value from \cite{Sing08b}. This results in
%a simple vertical shift of any model calculation.
%Thus, for the moment no direct global Na abundance information
%will be given. Below, however, we will show that some abundance
%information could be extracted from the observations.

\begin{figure}
\includegraphics[angle=0, width=0.48\textwidth]{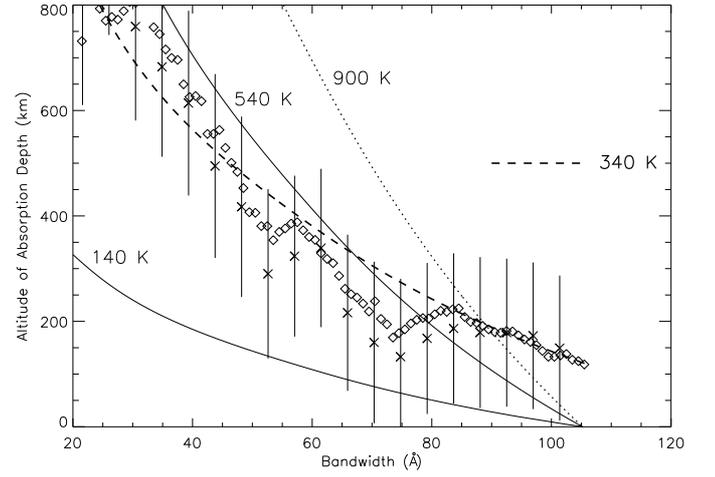}
%{Obs_vs_altitude_Var_T_Ref0.0149145_v22.eps}
\caption{Plot of the absorption altitude as a function of the
wavelength band. Measurements of Sing et~al. (2008a) are plotted:
photometric with crosses and spectroscopic with diamonds. The
typical error bars of all evaluations are shown only over the
photometric evaluations to avoid confusion. The dashed line shows
the best--adjusted \emph{IHUA} (isothermal hydrostatic and uniform
abundance) model to the lowest altitude observations,
corresponding to a temperature of 340\,K. For comparison the
theoretical \emph{IHUA} models are shown at $T=$~140\,K, 540\,K
(solid lines), shifted in ordinate toward the lower end of the
first error bar at 105\,\AA\ in order to show the constraints on
the model slopes (not affected by the applied shift) related to
the observational error bars. This leads to an estimate of the
temperature at the lower levels to be 340\,$\pm$\,200\,K. Similar
errors on the order of $\pm$\,200\,K are found for each
observational region of about one scale height (see text).
Finally, the 900\,K model (dotted line) corresponding to the
sodium condensation temperature (see Fig.~\ref{fig:TPprofile}) is
also shown to illustrate its incompatibility with the
observations.} \label{fig:compT}
\end{figure}

%\begin{figure}
%\includegraphics[angle=0,width=0.48\textwidth]{Obs_vs_altitude_Var_Tmin_bis_Ref0.0149145_v22.eps}
%\caption{Same as in Figure~\ref{fig:compT}. Here the \emph{IHUA}
%models shown are selected in order to show the most extreme
%possible temperatures one can use to reasonably represent the
%observed AD variations, within the shown error bars (see text).
%The lowest temperature in that region of the atmosphere should
%be~: 100K $<$ $T$ $<$ 500K.} \label{fig:Tmin}
%\end{figure}

Figure~\ref{fig:compT} shows the expected variations for fixed
abundances of NaI over the whole atmosphere, overplotted with
isothermal \emph{IHUA} models at different temperatures. As
explained above, the higher the temperature, the higher the scale
height in the atmosphere and thus the steeper the slope. Here, the
Na abundance corresponding to each temperature is assumed to be
constant, while the model representations in
Figure~\ref{fig:compT} are arbitrarily vertically shifted (thus
keeping the same slope at each given wavelength) in order to match
the observations above the reference level.

\begin{figure}
\includegraphics[angle=0,width=0.48\textwidth]{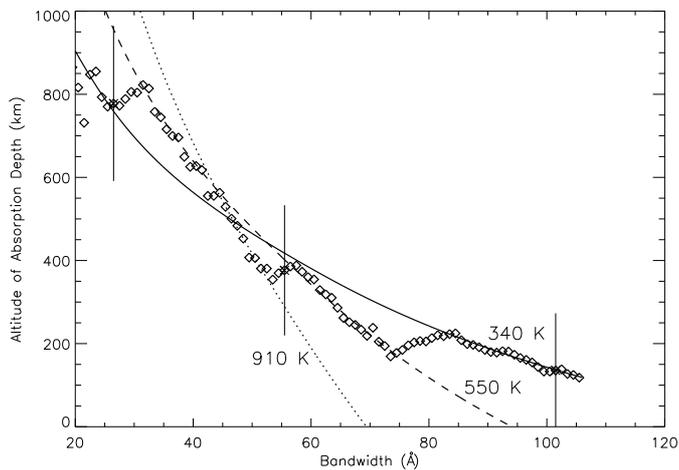}%{Comp_AD_T_NasH_1.5e-6_vs_low_Alt_v2.eps}
\caption{Same as in Figure~\ref{fig:compT} with \emph{IHUA} models
matching the observations up to 800~km in altitude. At the lowest
altitude, the temperature is found to be $\sim$~340~K (solid
line). At altitude above $\sim$~200~km, we identified layers with
different temperatures at about 550~K (dashed line), 910~K (dotted
line), then back to 550~K (dashed line). A few error bars are
indicated to underline that, in the lower altitude region [III]
from 0 to 800~km above the reference level, the observations are
on the average incompatible with an \emph{IHUA} model slope
corresponding to $\sim$900~K, the sodium condensation temperature,
demonstrating that condensation has to take place in the planetary
atmosphere.} \label{fig:evalTNasH}
\end{figure}

\section{Results and discussion}
\label{sec:discuss}

\subsection{The temperature profile at low altitudes (region [III])}
\label{T at lower altitudes}

Using the method described above, we start by evaluating the
temperature at altitudes just above the reference level (altitude
0--225~km, bandwidths 75--110~\AA ), with the reference level
corresponding to the level where the sodium abundance drops
sharply ($\sim$3~mbar level). The temperature of the atmosphere
just above the reference level (see Fig.~\ref{fig:compT}) is found
by the \emph{slope} of the observed variations (see
Eq.~\ref{T_dz}). In this atmospheric layer (region [III] as shown
in Fig.~\ref{fig:obs1}), the absorption depth varies slowly as a
function of the bandwidth. As a consequence, we find the lower
temperature close to $\sim$~340~K. Because this temperature
appears to be extremely low, the assumption of uniform abundance
is likely not to be valid for the scale height just above the
reference level, a region probably still affected by the
Na--depletion mechanism (see Sect.~\ref{The Na abundance
profile}). From Eq.~\ref{z_lambda}, the measured altitude is
determined by the quantity of $[Na/H]\times\sigma_{NaI}(\lambda)$.
If the Na abundance drops with increasing altitude, for a given
optical depth through the atmosphere, the Na cross section needs
to be larger to compensate, making the transmission spectrum
optically thick closer to the Na core. As a result, a depletion in
sodium causes a shallower slope (or even a plateau) in the
observed absorption depth versus bandwidth. In other words, if the
temperature is measured by the slope of the absorption altitude as
a function of wavelength assuming a constant abundance, a decrease
in sodium abundance with altitude implies underestimation of the
temperature (Eq.~\ref{Eq:T}). The temperature of 340~K is
therefore a lower limit of the temperature in the layer located in
the altitude range from 0 to 200~km above the reference level. In
this layer, the temperature is likely higher and the abundance of
sodium is very likely decreasing with altitude.

Apart from the shallow slope of absorption as a function of the
bandwidth described above, we observe two sodium absorption--depth
plateaus at 80\,\AA\ and 55\,\AA\ bandwidth. This can also be
interpreted as abundance drops at the corresponding altitudes. We
however ignore these two plateau regions because they are entirely
immersed within the error bars domain (see Fig.~\ref{fig:compT}).
We thus assume locally constant abundance (at least over one scale
height) to estimate the local temperature in the different layers
of the atmosphere.
%As the absorption depths measurements are obtained step by step in different
%bandwidths, these are not independent evaluations, we have thus to
%compare observations separated by at least about one scale height
%in order to reach independent evaluations.

%Over the 150-400~km altitude range (about one scale height),
%using the slope of absorption depth (Fig.~\ref{fig:compT}),
%we find a the temperature of 500$\pm250$K (1-$\sigma$ error bars).
%Indeed the corresponding model
%calculations pass along the lower parts of the error bars at
%larger bandwidths and the upper parts of the error bars at smaller
%bandwidths.

%As already mentioned the 250K temperature matches
%quite well the slope of the observations.

%As shown, the highest possible temperature still compatible with
%the extremes of the AD error bars is 500K, while very low
%temperatures as low as 100K are still possible~: 100K $<$ $T$ $<$
%500K.

%This way to evaluate the error bars on the different
%temperature evaluations was repeated at each following steps,
%and shows that over one scale height, which is our
%minimum size step, a typical error bar on the temperature will be
%stable and of the order of $\pm 250$K in particular because with
%increasing temperature (the following steps) the scale height
%increases and thus the domain of altitude over which the error
%estimate is made also increases.

\begin{figure}
\includegraphics[angle=0,width=0.48\textwidth]{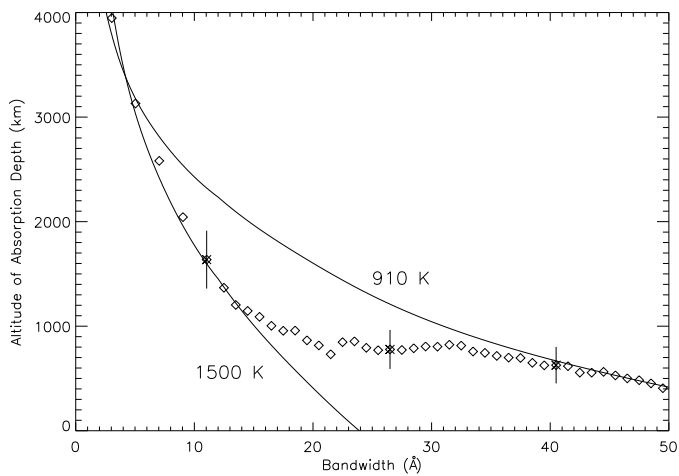}%{Comp_AD_higher_Alt_v3.eps}
\caption{Same as Figure~\ref{fig:evalTNasH} for narrower bandpass,
\emph{i.e.} corresponding to regions [I] and [II] at higher
altitudes from 800~km to 4\,000~km.} \label{fig:evalTNasHhigher}
\end{figure}

The temperature profile can thus be obtained step by step toward
increasing altitude by analyzing the observed AD variations in
narrower bandpasses. When starting from wavelength bandwidths of
110~\AA , working toward shorter bandwidths, and assuming constant
abundance, the AD variations can be interpreted in terms of
temperature changes as a function of altitude with the following
characteristics (see Fig.~\ref{fig:evalTNasH}): \\
i) a lower limit of 340~K in temperature is obtained at the
lowest altitudes, \\
ii) between altitudes of 200~km and 400~km,
the temperature is found to be about 550~K, \\
iii) the AD slope constraints the temperature at about 910~K in
the layer
between 400~km and 600~km altitude, \\
iv) the temperature seems to drop back close to 550~K between 600~km and 800~km altitude.\\
For each layer considered, the error bars on these temperature
evaluations are of about $\pm 200$~K, similar to the error
estimate made on the lowest temperature level at 340~K as shown in
Fig.~\ref{fig:compT}. Therefore, the observed temperature
variations correspond to a marginal rise in temperature with
altitude.

Most importantly, for about three scale heights from 200 to 800~km
above the altitude where the NaI abundance is seen to drop
sharply, the atmospheric temperatures are found to be within the
range $\sim$~550~--~910~K,  \emph{i.e.} below the sodium
condensation temperature (Figure~\ref{fig:TPprofile}). This result
favors the sodium condensation scenario to explain the sodium
abundance drop as described in (\cite{Sing08a, Sing08b}). Recent
global climate models of HD\,209458\,b also suggest cool
temperatures of $\sim$~800~K (at mbar pressures) are possible on
the night--side of the planet and terminator (\cite{Showman09}),
further supporting this sodium condensation model.

\begin{figure}
\includegraphics[angle=0,width=0.48\textwidth]{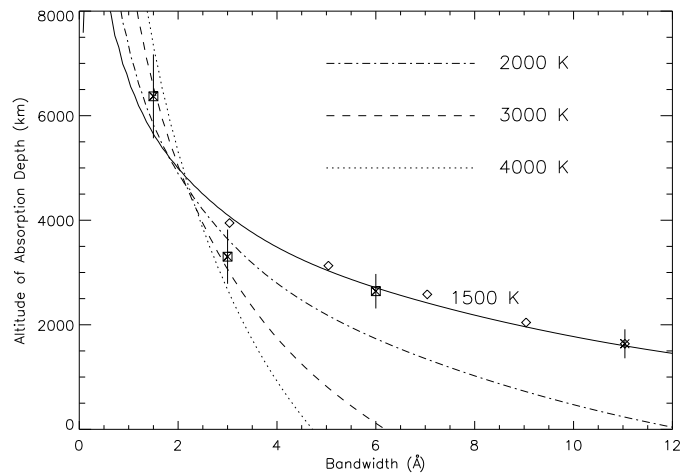}%{Comp_AD_Obs_T_Highest_Alt_Nash_1.5e-6_v2.eps}
\caption{Same as in Figure~\ref{fig:evalTNasH} for the narrowest
bandpass (region [I]). Near to the line center, the cross section
is the highest, therefore probing the highest altitudes. The
ground--based measurements obtained by Snellen et~al. (2008) are
plotted with crossed squares together with the spectroscopic
evaluations (diamonds) of Sing et~al. (2008a). The variations of
the absorption depth (AD) between 1.5 and 3\AA\ bandpass are
consistent with a temperature from 1\,500 to 4\,000~K with the
most plausible value on the order of 2\,500~K.}
\label{fig:evalTNasHhighest}
\end{figure}

\begin{figure*}
\includegraphics[angle=0,width=\textwidth]{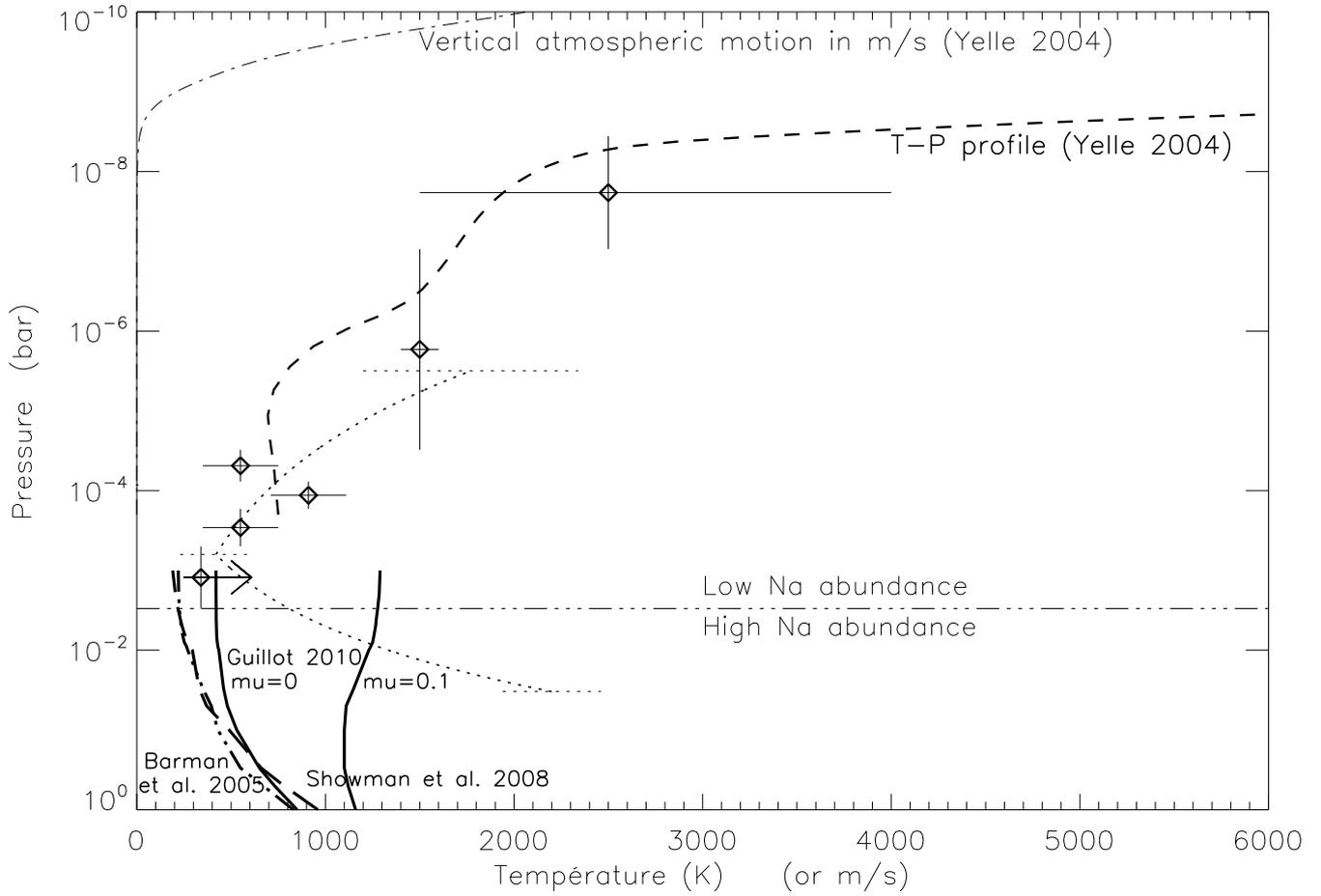}%{Fig9_new.ps}%{Pressure_T_Sol_1.5e-6_Log_Error_Yelle_e-9_NTmin.eps}
\caption{Plot of the ``integrated--averaged'' T--P profile
(diamonds) obtained in Sect.~\ref{The T--P profile}. The vertical
error bars show the pressure domain of the corresponding
temperature estimates. The triple--dot--dashed line shows the
altitude of the sodium abundance drop measured by
(\cite{Sing08a,Sing08b}). At pressure between $10^{-3}$ and
$10^{-5}$~bar, the low temperature ($T<800$K) favors the sodium
condensation scenario to explain the sodium abundance drop. Our
T--P profile is consistent with the T--P profile obtained by Sing
et~al. (2008b) using a parametric fit of the same data set (dotted
line) without the addition of the latest observations at higher
altitude levels from Snellen et~al. (2008). For comparison several
theoretical T--P profiles are over plotted. At higher pressures
(P$>$1~mbar), models of Barman et al. (2005; dash--triple--dotted
line), Showman et al. (2008; long dashed line) and Guillot (2010;
two solid lines, one for $\mu =0$ at the terminator limb and the
other at $\mu =0.1$ close to the terminator limb) are shown. At
lower pressure levels, the Yelle (2004) theoretical T--P profile
is also shown for comparison (dashed line). The dot--dashed line
shows the vertical velocity of the hydrodynamic outflow in the
Yelle (2004) model (in m/s). This velocity is negligible at
pressure above a few $10^{-8}$~bar where the hydrostatic
assumption is therefore justified.}

\label{fig:T-P}
\end{figure*}

\subsection{The temperature profile at higher altitudes(regions [I] and [II])}
\label{T at higher altitudes}

The temperature at higher altitudes can be estimated by looking
for the AD variations over narrowest bandpasses. There, an
important and significant leveling off (region [II]) of the AD
variation at about 800~km altitude, between 32\AA\ and 20\AA\
bandwidths, is seen (note in Fig.~\ref{fig:evalTNasHhigher} its
position relative to the observational error bars). This plateau
in absorption depth might be interpreted as a local drop in
temperature but this would correspond to an extremely low
temperature of less than 100~K ! This plateau is more likely
related to a decrease in sodium abundance which will be further
discussed in Sect.~\ref{The Na abundance profile}.

Above that long leveling off, \emph{i.e.} above 800~km altitude
(region [I]), the AD increases again rapidly toward shorter
bandwidths (from 15\AA\ bandpass and below), revealing there a
rise in temperature. Using an \emph{IHUA} model fit to the AD
profile, we find a typical temperature of about 1500~K from about
800~km up to 4\,000~km altitude (Fig.~\ref{fig:evalTNasHhigher}).
The measurement at 4\,000~km corresponds to the narrowest 3\AA\
bandpass in the very center core of the line in HST observations.
It is noteworthy that this 1\,500~K atmospheric layer is observed
to be isothermal over nearly 6 scale heights. Over this wide
altitude range, the error bar on the estimated temperature is
rather small ($\pm$100~K). There the temperature is tightly
constrained in the 1\,400\,K--1\,600\,K range.

After publication of the HST data analysis by Sing et~al. (2008a),
new measurements of the Na\,{\sc i} absorption depth have been
obtained by Snellen et~al. (2008) using ground--based
observations. These measurements have been obtained in narrower
bandpasses, providing larger absorption signatures, therefore
corresponding to absorptions at higher altitudes. These new
measurements are shown in Fig.~\ref{fig:evalTNasHhighest} together
with those of Sing et~al. (2008a). Although these various
measurements have been obtained using very different techniques,
and despite that ground--based measurements are known to be
extremely difficult because of the Earth atmospheric perturbation,
the very good agreement between the various data sets strengthens
the confidence in those measurements permitting us to extend our
analysis in altitude up to more than 6\,500~km.

On the contrary, we decided not to use the recent ground--based
observations of Langland--Shula et~al. (2009) because their result
suffers from uncertainties in the process of continuum fitting and
subtraction of the Earth atmosphere absorption.

We find that to match the variations of the absorption depth
toward the narrowest 1.5~\AA\ bandwidth as measured by Snellen
et~al. (2008) , the temperature at altitudes above 4\,000~km needs
to be on the order of 2\,500~K and within a 1\,500\,--\,4\,000~K
range (Fig.~\ref{fig:evalTNasHhighest}).

\subsection{The ``integrated--averaged'' T--P profile}
\label{The T--P profile}

An important comment here before presenting the result is to
recall that the evaluated T--P profile is extracted from
quantities ``integrated'' along the cords passing through portions
of the limb presenting gradients, {\it e.g.} from the night to day
sides, as well as ``averaged'' over the planetary limb, {\it e.g.}
from equatorial to polar regions. In a first step, such an
``integrated--averaged'' T--P profile could be compared to 1--D
model calculations, although it is clear that later steps, when
observations permit it, would be to compare them to more
sophisticated 3--D models as illustrated in {\it e.g.} the Iro
et~al. (2005) and the recent Fortney et~al. (2010) studies.

From the temperature--altitude relation obtained in Sects.~\ref{T
at lower altitudes} and~\ref{T at higher altitudes}, and assuming
hydrostatic equilibrium, we can calculate the
temperature--pressure (T--P) profile. We start by using a
reference pressure of 3~mbar at the reference 0~km altitude. The
pressure is calculated step by step, starting from the zero
altitude level. The result is shown in Fig.~\ref{fig:T-P}.

Acknowledging that the sodium condensation occurs at temperatures
below 800\,--\,900~K, the temperatures measured in the layer
between 0 and 800~km above the reference level are found to be low
enough to allow condensation. This condensation seems to take
place at pressures between 3~mbar and 10~$\mu$bar.

As already mentioned in the introduction, at pressures above
3~mbar, the evaluated observational profiles do not present
variations similar to the model predictions as calculated in the
literature specifically for HD\,209458\,b ({\it e.g.} Barman et
al. 2002; Burrows et al. 2003; Iro et al. 2005; Barman et al.
2005; Fortney et al. 2008; Showman et al. 2008; Showman et al.
2009; Guillot 2010; Fortney et al. 2010). Three of them (Barman et
al. 2005; Showman et al. 2008; Guillot 2010) are presented in the
lower part of Fig.~\ref{fig:T-P}. It is interesting to note that
all temperature predictions at about 3~mbar are in good agreement
with our evaluation simply made from the direct estimation of the
scale height at these levels. We also point out the large
variations revealed by the two near terminator--limb ({\it i.e.}
near the transit observed limb) model calculations from Guillot
(2010; one for $\mu =0$ at the terminator limb and the other at
$\mu =0.1$ close to the terminator limb) as also shown in
Fig.~\ref{fig:T-P}. This may indicate that indeed such model
predictions are difficult to compare to observations according to
their extreme sensitivity to the precise limb geometry and, as
already said, to the integrated nature of the observations along a
grazing line of sight. Also, one could argue that at higher
pressures, it is not surprising that observations (and their
interpretation) differ from theoretical models, since for
instance, a large heat advection could be critical in the
troposphere leading to strong departures from equilibrium
temperature profiles.

Above this condensation layer, a progressive rise in the
temperature with altitude is needed to explain the profile of the
sodium line toward narrower bandpasses. The temperature in the
upper atmosphere is found to reach $\sim$\,2\,500~K at the very
low pressures of $\sim$\,3 to $\sim$\,90~nbar. This is likely the
signature of the base of the thermosphere, which is the expected
transition between the cool lower atmosphere and the hot upper
atmosphere where hydrodynamic atmospheric escape takes place. This
scenario agrees with theoretical models of the atmospheric escape
as published by, \emph{e.g.}, Lecavelier des Etangs et al. (2004),
Yelle (2004) overplotted in Fig.~\ref{fig:T-P}, Tian et~al.
(2005), Yelle (2006), Mu\~noz (2007), and Murray-Clay et~al.
(2009).

A summary of our T--P profile is given in Table~\ref{table:1}. The
scale heights $H$ are given for each temperature layer. This shows
that the layers have thicknesses of at least one or more scale
heights. With a pressure change by a factor of 2.7 over one scale
height under hydrostatic equilibrium, Fig.~\ref{fig:T-P} shows
that the bottom layer at the lowest temperature of $>$~340~K has a
thickness of nearly 2~scale heights. The three following layers at
550~K and 910~K are over about 1 scale height, while the
atmospheric layer with a constant temperature at $\sim$\,1\,500~K
extends over nearly 6 scale heights. Finally the last upper layer
at about 2\,500~K extends over more than 3 scale heights.

The calculated T--P profile is valid when assuming local
hydrostatic equilibrium. This assumption could be invalid because
of the hydrodynamical escape of the upper atmosphere.
Figure~\ref{fig:T-P} shows that the theoretical escape velocities
triggered by the hydrodynamic flow as evaluated by Yelle (2004)
are negligible below the altitude corresponding to a pressure of
$\sim$\,10$^{-9}$~bar. This shows that the local hydrostatic
equilibrium assumption is valid over our studied range, even at
the highest altitudes where we estimate a temperature of about
2\,500~K.

%_____________________________________________________________
%
\begin{table}
\caption{Evaluated atmospheric parameters}             % title of Table
\label{table:1}      % is used to refer this table in the text
\centering                          % used for centering table
\begin{tabular}{c c c c c c c}        % centered columns (4 columns)
\hline\hline                 % inserts double horizontal lines
$z$~$_{\rm min}$ & $z$~$_{\rm max}$ &  $T\pm\Delta T$ & $H$ & $H/\Delta z$ & $P_{\rm max}$ & $P_{\rm min}$  \\    % table heading
(km) & (km) & (K) & (km) &  & (bar) & (bar)  \\    % table heading
\hline                        % inserts single horizontal line
    0 &  225 &    340$^{+200}_{~-~0}$ &  120 & 1.9 & 3.0e-3 & 5.0e-4 \\      % inserting body of the table
  170 &  390 &    550$\pm$200 &  200 & 1.1 & 5.0e-4 & 1.7e-4 \\
  350 &  620 &    910$\pm$200 &  340 & 0.8 & 1.7e-4 & 7.7e-5 \\
  620 &  800 &    550$\pm$200 &  200 & 0.9 & 7.7e-5 & 3.1e-5 \\
  800 & 4\,000 &   1\,500$\pm$100 &  540 & 5.9 & 3.1e-5 & 9.4e-8 \\
 3\,500 & 6\,500 &   2\,500$^{+1\,500}_{-1\,000}$ &  920 & 3.3 & 9.4e-8 & 3.6e-9 \\
\hline                                   %inserts single line
\end{tabular}
\end{table}
%
%_____________________________________________________________
%

\subsection{The Na abundance profile}
\label{The Na abundance profile}

The sodium abundance [Na/H] cannot be constrained in absolute
terms according to the mentioned degeneracy between abundance and
pressure in the absorption spectrum (see Lecavelier Des Etangs
et~al. 2008a and Eq.~\ref{z_lambda}). However, here the degeneracy
can be solved by assuming a reference pressure at the  reference
level (see Section~\ref{Ref_Pressure}). Indeed we used a reference
pressure of 3~mbar for the altitude corresponding to our reference
level at 1.49145\% absorption depth. Using in addition the sodium
abundance below the reference level as evaluated by Sing et~al.
(2008b) in the lower layers of the atmosphere,
[Na/H]$\sim$3$\times$10$^{-6}$ (solar abundance equals
1.5$\times$10$^{-6}$, Asplund et~al. 2006), we can deduce the
sodium abundance evolution in the upper layers. In agreement with
the condensation scenario, the sodium abundance drops just above
the reference level and then, following the upward evaluation of
Sing et~al. (2008b), it should be reduced on the average over all
the upper layers by about a factor 10.

Starting from the value at the reference level, the Na abundance
is evaluated at higher levels using a step--by--step procedure
with \emph{local IHUA} models. A constant [Na/H] abundance in the
isothermal layers obtained in Sects.~\ref{T at lower altitudes}
and~\ref{T at higher altitudes}, has to be present, in agreement
with our \emph{IHUA} hypothesis.

\begin{figure}
\includegraphics[angle=0,width=0.48\textwidth]{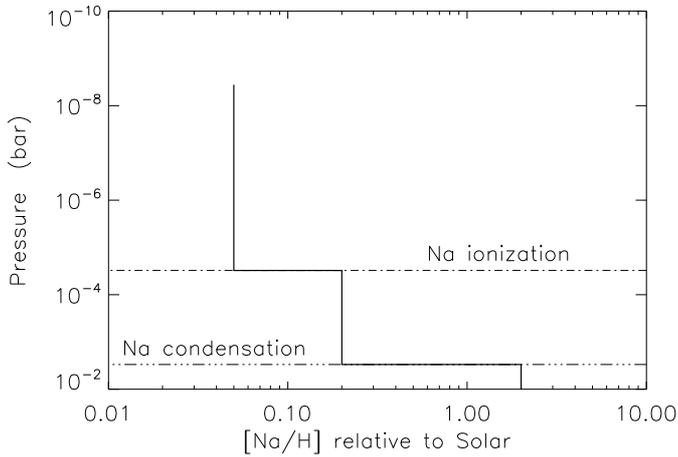}%{Pressure_NasH_MAx_Sol_1.5e-6_Log_Error_1_2_N.eps}
\caption{Summary of the Na abundance estimates using the
\emph{IHUA} model. Because our approach is local we can only
identify here the decreases in sodium abundance by a factor of
about 4 at $3\times 10^{-5}$~bar (dot-dashed line), although NLTE
effects could perturb that estimation (see text). The drop in
abundance by about a factor of 10 showed at the condensation level
(at 3~mbar, tripled--dot--dashed line) is taken from Sing et~al.
(2008a,b).} \label{fig:T-Max-NasH-double}
\end{figure}

A plateau in the measured absorption depth is observed at an
altitude of about 800~km. This plateau is a kind of anomaly: with
the increase of absorption cross section in the core of the line,
the AD should rise toward smaller bandwidths. If the plateau was
extending to the smallest bandwidths at the center of the line,
this would reveal the absence of the absorber at altitudes above
this plateau. However, here the plateau has a limited extension
from $\Delta\lambda_1\sim 15$~\AA\ to $\Delta\lambda_2\sim
30$~\AA\ bandwidth. The absorption depth rises clearly again at
bandwidths smaller than about 15~\AA\ (see
Fig.~\ref{fig:evalTNasHhigher}). This reveals a drop in the sodium
abundance at the altitude of the plateau, but the sodium is still
present above, thereby producing the observed absorption rise
toward the line center. Along the plateau, the variation on the
line cross section with wavelength, from $\sigma(\lambda_1)$ to
$\sigma(\lambda_2)$, is compensated for by the variation in the
sodium abundance, ${\rm [Na/H]}_1$ and ${\rm [Na/H]}_2$ above and
below the plateau altitude. Using Eq.~\ref{z_lambda}, the altitude
of the plateau $z(\lambda)$ being constant from $\lambda_1$ to
$\lambda_2$, we find that
\begin{equation}
\left(\frac{ {\rm [Na/H]}_1 } { {\rm [Na/H]}_2}\right)
=
\left(\frac{\sigma(\lambda_2)} {\sigma(\lambda_1)}\right).
\end{equation}
Because here the absorption is dominated by the damping wings of
the sodium line, we have
\begin{equation}
\left(\frac{\sigma(\lambda_2)} {\sigma(\lambda_1)}\right)
\cong
\left(\frac{\lambda_2-\lambda_0} {\lambda_1-\lambda_0}\right)^{-2}.
\end{equation}
With $\Delta\lambda_2/\Delta\lambda_1\sim$2~ for the plateau at
800~km of altitude, we can determine that this plateau may be
explained by a drop in sodium abundance by a factor
[Na/H]$_2$/[Na/H]$_1$~$\sim$~4. According to our approach, which
is essentially local, we can conclude about this relative local
drop, while as stated earlier, absolute values of the sodium
abundances could not be assessed. To represent these absolute
values, we used the result of the global approach of Sing et~al.
(2008a,b) as shown in Fig.~\ref{fig:T-Max-NasH-double}.

Considering the sodium abundance drop evaluated by Sing et~al.
(2008a,b) at the condensation level to be on the order of a factor
of 10, we assumed that [Na/H]$_2$ should be on the order of 0.2
solar, while the Na abundance above the altitude of the plateau
should be [Na/H]$_1$$\sim 0.05$ solar.
Figure~\ref{fig:T-Max-NasH-double} shows the Na abundance
variations obtained from the observed AD over the whole domain
sampled in Fig.~\ref{fig:T-P}.

To match the observed AD plateau, one has to assume a local
reduction in the Na abundance by a factor of about~4, which could
indicate of a possible physical effect in the atmosphere, such as
the ionization of NaI into NaII as predicted by several models,
\emph{e.g.} Fortney et~al. (2003). The conclusion on the drop in
the Na abundance to interpret the AD plateau is expressed in terms
of relative abundance, so this conclusion does not depend upon any
assumption about pressure or absolute abundances.

We are evaluating variations in the Na abundance in low volume
density regions where nonlocal thermodynamic effects (NLTE) should
take place, {\it i.e.} below 1~mbar as evaluated by Barman et~al.
(2002). There radiative processes dominate collisional ones and
the NaI population levels could be significantly away from the
assumed local thermodynamic equilibrium (LTE) ones. As long as
such effects do not vary significantly over one scale height, our
temperature evaluations should not be affected by these NLTE
effects. On the contrary, such effects certainly should perturb
abundance evaluations, and in particular, the previously mentioned
factor 4 drop of abundance at $3\times 10^{-5}$~bar could as well
be the signature of such NLTE effects at these altitudes, leading
probably to an underestimation of the evaluated drop (Barman
et~al. 2002). If the corresponding signature is indeed due to Na
ionization, it is in reasonably good agreement with the Barman
et~al. (2002) evaluation who found that Na would be mostly ionized
in the limb for P $<$ 10$^{-6}$ bar and only partially ionized
($\sim$~3\%) below this point.

\section{Conclusion}
%====================
\label{sec:conclusion}

Using the measured NaI absorption depth as a function of the
bandwidths, we estimated temperature variations as a function of
altitude. Our temperature profile is inconsistent with one of the
two scenarios of Sing et~al. (2008b) (scenario with constant
temperature and ionization at low altitude).

We found layers at temperatures clearly below the NaI condensation
temperature showing that sodium condensation has to take place at
these levels, followed by a temperature rise at higher altitudes
up to about 2\,500~K. This rise is likely the signature of the
thermosphere needed to link the cool lower to the hot upper
atmosphere, as suggested by the observations of atmospheric escape
(\cite{Madjar03,Madjar04,Linsky10}), observations of high
temperatures (\cite{Ballester07}), and models of the escape
mechanism
(\cite{Lecavelier04,Yelle04,Tian05,Yelle06,Munoz07,Murray09}).

The results of our analysis favor the sodium condensation scenario
to explain the deficiency of sodium observed in the line core, and
the possible presence of an Na ionization layer just below the
base of the thermosphere. Our temperature profile presents the
following patterns:

-- a rise from 340~K to 1\,500~K over $\sim$800~km (about 5~scale
heights) in the pressure range of $3\times 10^{-3}$--$3\times
10^{-5}$~bar;

-- a thick layer with a nearly constant temperature at about
1\,500~K. This layer has a thickness of about 3\,200~km
corresponding to nearly 6~scale heights for a pressure range of
$3\times 10^{-5}$--10$^{-7}$~bar;

-- a further temperature rise is observed at higher altitudes, up
to 2\,500~K at the highest altitudes about 6\,500~km above the
reference level, corresponding to a pressure of about $3\times
10^{-9}$~bar. This high temperature is likely related to the
bottom part of thermosphere whose presence is predicted by
theoretical models of the atmosphere.

-- an NaI abundance drop by a factor of~4. This is possibly due to
ionization at the $3\times 10^{-5}$ bar level, the strength of the
observed drop depending upon the importance of the NLTE effects at
these altitudes.

\begin{acknowledgements}
D.E. acknowledges financial support from the Centre National
d'Etudes Spatiales (CNES). D.K.S. was supported by CNES at the
start of this study. This work is based on observations with the
NASA/ESA Hubble Space Telescope, obtained at the Space Telescope
Science Institute (STScI) operated by AURA, Inc.

\end{acknowledgements}

%\bibliographystyle{aa} % style aa.bst
%\bibliography{Alfred} % your references in Alfred.bib, uses natbib

\end{document}